# Coulomb correlations evidenced in the magneto-optical spectra of charged excitons in semiconductor quantum dots

D. Y. Oberli, M. Byszewski, B. Chalupar, E. Pelucchi*, A. Rudra and E. Kapon

Laboratory of Physics of Nanostructures, Ecole Polytechnique Fédérale de Lausanne

CH-1015 Lausanne, Switzerland

Abstract:

The emission spectral pattern of a charged exciton in a semiconductor quantum dot is composed of a quadruplet of linearly polarized lines in the presence of a magnetic field oriented perpendicularly to the direction of the photon momentum. By measuring the Zeeman splittings, we obtain the effective g factors of the carriers and find that the hole g factor is very sensitive to the QD shape asymmetry. By comparing the effective g factors obtained for the neutral and the charged excitons in the same quantum dot, we uncover the role of Coulomb correlations in these excitonic states.

Spin degree of freedom of carriers in semiconductor quantum dots (QDs) could serve as quantum bits to store information in spin-based devices[1]. A key property of an electron (or hole) confined to a QD is the effective g factor that measures the Zeeman splitting of the ground state in an applied magnetic field and depends on the orientation of the magnetic field. To manipulate coherently the spin of a carrier in a QD, it is necessary to understand the anisotropy of these splittings and their sensitivity to the dot size and shape[2-3]. Zeeman splittings in QDs have been investigated by capacitance spectroscopy[4], by magneto-photoluminescence[5] and by transient nonlinear optical techniques[6]. While transport techniques probe the g-factors of electronic states, optical techniques probe the Zeeman splittings of excitonic states. In principle, probing different states could yield different values if the Coulomb interaction between several carriers alters the many-body state of the confined carriers. A separate determination of carrier g-factors from excitonic states was realized in a few specific cases, either by using an asymmetric QD or by applying a transverse magnetic field without addressing this issue (see ref. 5). The anisotropy of the effective g factors on the orientation of the magnetic field was recently studied on ensemble measurements of self-assembled quantum dots (SAQDs)[7-8], however, the sensitivity of the g factors on the QD asymmetry could not be determined.

In this study, we present a novel experimental approach to determine separately the effective g factors of an electron and a hole in individual QDs. When a QD is placed in an external magnetic field, the emission spectral pattern of a charged exciton is composed of a quadruplet of lines if the field direction is perpendicular to the photon momentum (Voigt configuration). We demonstrate how the effective g factors of the carriers can be measured from the Zeeman splittings of this quadruplet and explain the polarization selection rules. By comparing the emission from the neutral and charged excitons in the same QD, we show that Coulomb correlation modifies significantly the effective g factors.

We use pyramidal $In_{0.1}Ga_{0.9}As/Al_{0.3}Ga_{0.7}As$ QDs that allow us to observe simultaneously the positively-charged and the negatively-charged excitons in the same photoluminescence (PL) spectrum. The measurements were performed on individual QDs at a temperature of 10 K using a microphotoluminescence setup[9]. The cryostat containing the sample was inserted in the room temperature bore of a superconducting magnet, which generates magnetic fields up to 6.5T. The PL was excited with a continuous-wave Ti:sapphire laser operating at 700 nm and dispersed in a spectrograph equipped with a Si charged-coupled device detector. The spectral resolution was 40 μeV and the spectral precision was ±5 μeV by fitting the spectral lineshape. We found that the linewidth of an excitonic transition lay between 80 and 110 μeV for the investigated QDs.

A typical PL spectrum of a pyramidal QD at zero magnetic field is presented in Fig.1a. It consists of a series of four major lines, which correspond to the recombination of the negatively-charged exciton ($X^-$) of a biexciton (2X), of a neutral exciton (X) and of the positively-charged exciton ($X^+$) ordered by increasing energies[10]. The linearly polarized emission spectra evidence the existence of a doublet for both the neutral exciton and the biexciton while the charged exciton lines do not feature any resolvable splittings. Consequently, we attribute the splitting of X (2X) to the anisotropic part of the exchange interaction, which has been thoroughly studied in SAQDs (Ref. 5). The weaker lines in the PL spectra have been previously identified as emission from excitonic complexes wherein one of the holes occupies an excited state of the QD[11].

The application of a magnetic field in the Voigt configuration leads to striking modifications of the PL spectra of the QD as shown in Fig. 1b, where we display the linear polarization spectra of the QD under an applied magnetic field of B = 6.5 T. The emission is linearly polarized in a direction that is either parallel or perpendicular to the field direction. For a given orientation of the polarization, the emission of the charged excitons consists of a doublet of lines of nearly equal intensities whereas that of the neutral exciton (biexciton) consists of a doublet with a weak line on the low (respectively high) energy side of a dominant line. This latter emission pattern is the well-known signature[12] of the hybridization between the "dark" exciton states of total angular momentum ($m_z=\pm 2$) and of the "bright" exciton states of angular momentum ($m_z=\pm 1$) that is caused by a magnetic field when applied in a direction perpendicular to the z axis of a pyramidal QD (the orientation of **B** is sketched in inset of Fig. 1a). In the inset of Fig.1b, we display the measured energy splitting of the neutral exciton doublet versus the magnetic field for each polarization of the emission. It increases nearly quadratically with the magnetic field as expected (the fitting will be described later). The magnetic field dependence of these splittings is determined by the Zeeman interaction between the spin of the carriers and B and by the electron-hole (e-h) exchange interaction as we will explain later.

We will first focus on the emission patterns of the charged excitons. In striking contrast to the neutral exciton behavior in the magnetic field, we observe that the doublet components have nearly equal intensities at all values of the magnetic field: at B = 6.5T, e.g., the spectra of the negatively-charged exciton are expanded in Fig. 2. Moreover, we find that the splitting of each charged exciton doublet *increases linearly with B* as shown in the inset for the case of X⁻. These splittings were determined very precisely (error equal to ± 5 μeV) on the basis of a fit of the doublets to Gaussian lineshapes including a spectral background that varies linearly with photon energy.

The behavior of the charged exciton states in a magnetic field, their quadruplet emission pattern and their polarization properties, are fully determined by the Zeeman interaction alone. In a singly-charged excitonic complex, the e-h exchange interaction does not contribute to the splitting irrespectively of the QD symmetry because of the Kramers degeneracy of the charged exciton states[13]. To explain the splitting pattern of the charged exciton emission, we consider that the quantum dot possesses a $C_{3v}$ symmetry along the growth axis of the pyramidal structure (z // [111]). The direction of the detected photons is chosen to coincide with this symmetry axis. The Zeeman Hamiltonian of an electron (or a hole) is given by the general expression: $H_Z^c = \pm \frac{1}{2} \sum_{i=x,y,z} g_i^c \mu_B \sigma_i B_i$, where c stands for the carrier type (electron or hole), $\mu_B$ is the Bohr magneton, $\sigma_i$ are the Pauli spin matrices and $g_i^c$ is a diagonal tensor representing the effective g factors of a carrier (minus sign applies to the hole Hamiltonian following the usual sign convention). It is important to note that, for $C_{3v}$ symmetry, $g_x^c = g_y^c$. The expression for holes supposes that the dominant contribution to the Zeeman interaction arises from the light-hole component of the ground state. It can then be described by a pseudospin ($\pm \frac{1}{2}$). This description is entirely validated by our previous experimental work[14], which proves that the hole ground state has a small but significant light-hole component. For a magnetic field oriented perpendicularly to the growth axis (taken as the x axis) the Zeeman Hamiltonian is non-diagonal in the spin basis $|z, \pm \frac{1}{2}\rangle$ where z is the quantization axis of the electron (hole) total angular momentum. The splitting pattern among the allowed optical transitions results directly from the diagonalization of the Zeeman Hamiltonian. The spin eigenvectors for each carrier type assume the form $\frac{1}{\sqrt{2}} \left[ |z, \frac{1}{2}\rangle \pm |z, -\frac{1}{2}\rangle \right]$ and the Zeeman energies are given by $\pm \frac{1}{2} g_x^c \mu_B B_x$. From these eigenstates we can simply derive the polarization selection rules in the dipole approximation for a photon emitted along z: four linearly polarized optical transitions with equal strength are predicted, two of them being polarized perpendicularly to the magnetic field direction and split by $\Delta E_\perp = \mu_B |g_x^e + g_x^h| B_x$ and the other two being polarized parallel to B and split by $\Delta E_{//} = \mu_B |g_x^e - g_x^h| B_x$. The polarization rules and the linear dependence of the splitting with B are perfectly confirmed by our experimental data on this QD, as shown in Fig. 2 for the case of X⁻, and also on other QDs.

From the measurement of the Zeeman splittings at B = 6.5 T we directly determine the modulus of the effective g factor for each carrier type. The data in Fig. 2 give $\left|g_x^e - g_x^h\right|$ and $\left|g_x^e + g_x^h\right|$ which, when combined, yield $\left|g_x^e\right| = (0.15 \pm 0.01)$ and $\left|g_x^h\right| = (0.42 \pm 0.01)$ in the case of X⁻. The attribution of the value 0.15 to the electron g factor is fully consistent with the determination made in a set of ten other QDs (see results in Fig. 3). An important point is that the signs of the g factors cannot be determined from measurements in the Voigt configuration. Nevertheless, the relative signs of the g factors can always be specified. For the particular QD of Fig. 1, we find that the signs are identical since the largest Zeeman splitting is obtained for a linear polarization that is perpendicular to B.

Since we observe both the emission of X⁺ and X⁻ in the same dot we can test whether the presence of an additional carrier in the dot modifies the value of the effective g-factors. We find indeed that the Zeeman splitting of X⁺ is smaller than that of X⁻ in each of the two polarizations. This means that the effective g factors differ depending on the sign of the charged exciton. Following the analysis done for X⁻, we obtain from the spectra displayed in Fig. 1 that, for X⁺, $\left|g_x^e\right| = (0.14 \pm 0.01)$ and $\left|g_x^h\right| = (0.38 \pm 0.01)$. Although the differences between the two sets of carrier g factors are small, the measured Zeeman splittings for X⁺ differ significantly from those of X⁻ as shown in Fig. 1b. These observations were confirmed in the magneto-optical spectra of ten other pyramidal QDs selected from the same sample. In Fig. 3, we report the carrier g factors, which are measured in ten other QDs for both X⁺ and X⁻. For the electron g factor, we observe a marked difference between the values obtained from the splittings of X⁺ and X⁻. The modulus of the electron g factor of X⁻ is always larger than that of X⁺: the mean value being $\left\langle\left|g_x^e\right|\right\rangle = 0.18$ for X⁻ and $\left\langle\left|g_x^e\right|\right\rangle = 0.12$ for X⁺; the standard deviation being σ = 0.02. This suggests that the presence of an additional carrier alters the single-particle electron wave function and, thus, modifies the electron g factor measured from either $X^+$ or $X^-$. In the case of holes, the effective g factor measured from $X^-$ can be either smaller or larger than that of $X^+$ (Fig. 3). Similarly, this evidences a distortion of the hole wave function in each QD. Additionaly, we find a broad distribution of effective g factors for holes, which contrasts with the narrow distribution found for electrons. Given that the error on the g factor is small, we infer from the distribution width that the hole g factor depends sensitively on the QD by a change of its shape or its volume. A change of QD volume would result in a correlation between the emission energy of the exciton and the hole g factor. However, no such correlation was found. On the other hand, a breaking of the axial symmetry of the QD will alter its shape and give rise to an anisotropic e-h exchange interaction. This interaction is known[15]-[16] to be a sensitive probe of a QD asymmetry and it is evidenced by the splitting of the neutral exciton emission into a doublet with linearly-polarized components. Experimentally, we find indeed a strong correlation between the magnitude of this splitting at zero magnetic field and the modulus of the hole g factor, as shown in Fig. 3 inset. The larger the hole g factor is, the larger the anisotropic e-h exchange splitting is. This behaviour confirms that the broad distribution of hole g factors results from a shape modification of the QD. This deduction is corroborated by the theoretical calculations of Pryor and Flatté[17], who predicted a large increase of the in-plane hole g factor of the dot shape resulting from a breaking of axial symmetry. Moreover, our observation of a narrow distribution of the electron g factors is in good agreement with their prediction. The insensitivity of the electron g-factor to the dot shape anisotropy was also confirmed theoretically by Shen et al.[18].

The sign of the electron g factor was specified only recently as being negative in InGaAs/GaAs-SAQDs using the dynamic nuclear polarization technique[19]. In general, magneto-PL experiments performed in the Voigt configuration yield the modulus of the carrier g factors but not the sign. The attribution of a negative sign to the electron g factor in the plane perpendicular to the growth axis (see Fig. 3) is then a choice for sake of consistency. Indeed, by performing experiments in the Faraday configuration in similar pyramidal QDs, we could determine that $g_z^e$ is negative when **B** is oriented along the [111] growth axis[20]. To our knowledge, a direct determination of the sign of the in-plane electron g factor has not been realized yet in InGaAs-SAQDs. We wish to emphasize that the sign of hole g factors in Fig. 3 is thus specified only relatively to that of the electron g factor. In these pyramidal QDs, we find a surprising result: the in-plane electron and hole g factors can have the same sign or an opposite sign (see Fig. 3). At present, an explanation of this effect is missing. A possible influence of the penetration of the hole wave function in the AlGaAs barrier might contribute to the sign of $g_h$ since in bulk GaAs and AlGaAs $g_h$ is positive.

The influence of an extra carrier in a QD on the modulus of the carrier effective g factors is a consequence of the Coulomb interaction between carriers. As shown in Figure 1 the Zeeman splitting for a given orientation of the polarization depends on the sign of the charged exciton. We find that the Zeeman splittings of $X^-$ are always larger than those of $X^+$ for all the investigated QDs. In contrast, we measured identical Zeeman splittings in the Faraday configuration for both $X^-$, $X^+$ and X when **B** was oriented parallel to the growth axis (Ref. 19). We infer from the sensitivity of the Zeeman splittings to the sign of the charged exciton that there is a significant degree of Coulomb correlation between the three carriers composing these excitonic complexes. In the simplest picture, the repulsion between electrons in the negatively-charged exciton could cause an additional spread of the wave function into the barrier and, thus, lead to a modification of the electron effective g factor.

In order to corroborate this interpretation, we analyze the Zeeman splittings of the neutral exciton measured in the Voigt configuration. In the inset of Fig. 1 we observe a quadratic dependence of the Zeeman splittings on the strength of B as previously described.

This dependence with B is well accounted for by adding to the Zeeman Hamiltonian the e-h exchange interaction for $C_{2v}$ symmetry written in the form[21]:

$$H_{exch} = \frac{\delta_o}{2} \sigma_z^e \cdot \sigma_z^h + \delta^* \sigma_x^e \cdot \sigma_x^h + \delta \sigma_y^e \cdot \sigma_y^h$$

where $\sigma_i$ are the Pauli spin matrices for an electron or for a hole (taken as a pseudo-spin ½ as described above). The first term in this expression is the isotropic exchange interaction that splits the neutral exciton into a doubly-degenerate "bright" state and a doubly-degenerate "dark" state, the second and third terms are the anisotropic exchange interactions that lift the degeneracy of these states without admixing them. Although the pyramidal QDs have $C_{3v}$ symmetry, we did not need to include additional terms in the Hamiltonian to describe our experimental results since a mixing of the bright and dark states was not evidenced in the PL in the absence of a transverse magnetic field. The energy splittings of the doublets can be derived by direct diagonalization of the spin Hamiltonian of an electron-hole pair, $H_s = H_z^e + H_z^h + H_{exch}$, and their expressions are given by

$$\Delta E_{//} = \sqrt{(\delta_o - 2\delta)^2 + (g_e^x - g_h^x)^2 \mu_B^2 B_x^2} \quad \text{and} \quad \Delta E_\perp = \sqrt{(\delta_o + 2\delta)^2 + (g_e^x + g_h^x)^2 \mu_B^2 B_x^2}$$

corresponding respectively to a direction of the linear polarization set either parallel or perpendicular to the applied magnetic field (here, **B** // **x**). We determine the effective carrier g factors for the neutral excitons by fitting the measured energy splittings with these expressions. From the fits shown in Fig. 1 inset, we obtain the values $|g_x^e + g_x^h| = 0.582 \pm 0.003$, $|g_x^e - g_x^h| = 0.17 \pm 0.01$, $\delta_o = 180 \,\mu eV$ and $\delta = 15 \,\mu eV$; from which we calculate $|g_x^e| = (0.206 \pm 0.006)$ and $|g_x^h| = (0.376 \pm 0.006)$. Experimentally, the neutral exciton emits a doublet of lines of equal intensities, with a splitting at zero magnetic field equal to $2(\delta^* - \delta) = 55 \,\mu eV$, which yields $\delta^* = (42 \pm 5)\mu eV$. By measuring the e-h exchange energy in other dots emitting closely in energy (± 1 meV), we find a small dispersion of the isotropic exchange energy around an average value of (196±10) μeV whereas the anisotropic exchange energy, $\delta$, varies over the range of values [0, 15 μeV]. The observed constancy of the isotropic part of the e-h exchange energy is understood because it is proportional to the probability of the electron and the hole to be at the same site[22], this probability remaining constant if the emission energy is unchanged. As the anisotropic part of the e-h exchange interaction is a sensitive probe of the QD asymmetry, it is then expected that a distortion of the QD shape away from cylindrical symmetry will increase the value of $\delta$ and $\delta^*$ as evidenced experimentally in the inset of Fig. 3. Our measurements are in agreement with the theoretical analysis of the excitonic fine structure developed for $In_{1-x}Ga_xAs/GaAs$ SAQDs[23]. Nevertheless, we notice that the experimental splitting energy between the "dark" excitonic states, given by $2(\delta^* + \delta) = 115\,\mu eV$, takes a value much bigger than the one calculated theoretically in Ref. 23, yet the other e-h exchange energies, $\delta_o$ and $2(\delta^* - \delta)$, are comparable.

The comparison of the electron effective g factors measured for X and X⁻ in the same QD reveals a significant difference, which is much larger than the estimated experimental uncertainty. The effective g factor is in a *simplified picture* a weighted average between the g factors in the barrier and in the dot materials. It is then affected by a tiny modification of the wave function penetration into the barrier, which can take place due to the Coulomb repulsion between the two electrons. As a result, one would expect the electron effective g factor in the dot to tend to its value in the barrier material. Because $g_e$ is positive in the barrier made of $Al_{0.7}Ga_{0.3}As$, the observed decrease of the modulus of $g_e$, when measured for X⁻, is consistent with our initial choice of a negative sign for the electron g factor. We also note that the relative change of the hole effective g factor measured for X and X⁻ is significant enough. A detailed explanation of this change would, however, go beyond a simple distortion of the hole wave function as the one described for the electron. We suggest instead that a mixing between the ground state and the excited states of the hole is taking place, which leads to a reinforcement of the light-hole component in the ground state wave function and, then, to an increase of the hole effective g factor[24]. We believe that our results will spur further theoretical works in order to investigate the interplay between Coulomb correlations[25] and the Zeeman interaction in semiconductor QDs of different shapes and degree of confinement[26]-[27].

In summary, we presented a novel and simple approach to investigate the carrier effective g factors in pyramidal QDs that relied on the Zeeman energy splittings of the charged excitons in a Voigt configuration. We observe that the effective g factors measured in this way depend on the sign of the charged exciton state, an effect that had not been previously evidenced in other types of QDs. We suggest that this effect originates from the Coulomb correlations between the carriers confined in the QD.

Figure captions:

**Fig. 1:** (color online) Polarized photoluminescence spectra of a pyramidal quantum dot measured (a) without and (b) with a magnetic field of 6.5 T in the Voigt configuration. The emission is analyzed into its linear polarization in directions parallel and perpendicular to the magnetic field. Upper inset: Zeeman energy splitting between the bright and dark excitons for each of the polarizations. Lines are fits using expressions given in the text. Lower inset: Sketch of pyramidal QD showing the photon emission direction (**k**) and the orientation of **B**.

**Fig. 2:** (color online) Emission spectra of the negatively-charged exciton (X⁻) at a magnetic field of 6.5 T displaying a doublet of transitions with the linear polarization set parallel or perpendicular to the field. Lines correspond to fits of the data with gaussians of identical width.

**Fig. 3:** (color online) Carrier g-factors of the (negatively-) positively-charged exciton measured for a set of ten pyramidal quantum dots. Inset: modulus of hole effective g factor plotted versus anisotropic exchange energy corresponding to $2(\delta^* - \delta)$.

* Present address: Tyndall National Institute, 'Lee Maltings', Prospect Row, Cork, Ireland

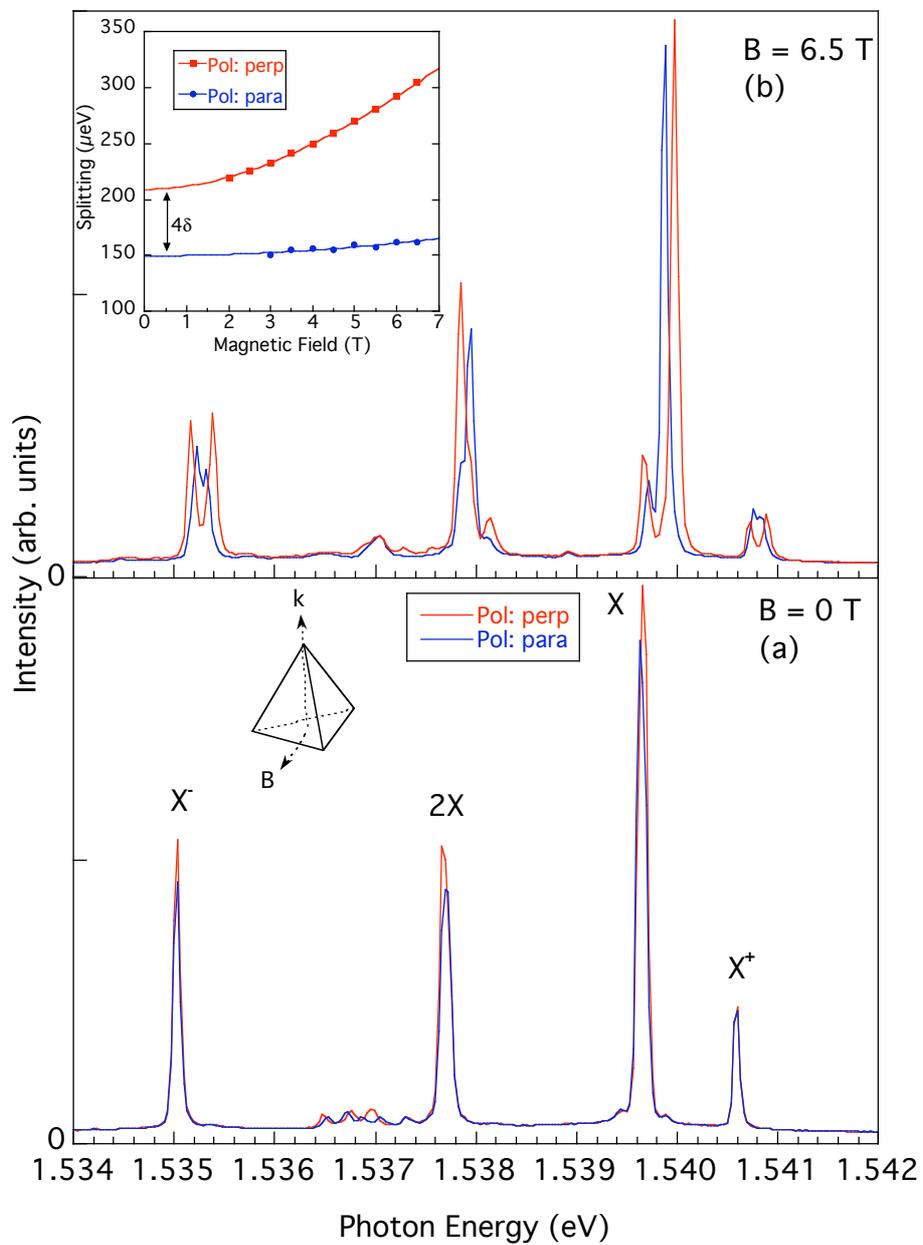

Figure 1

D.Y. Oberli

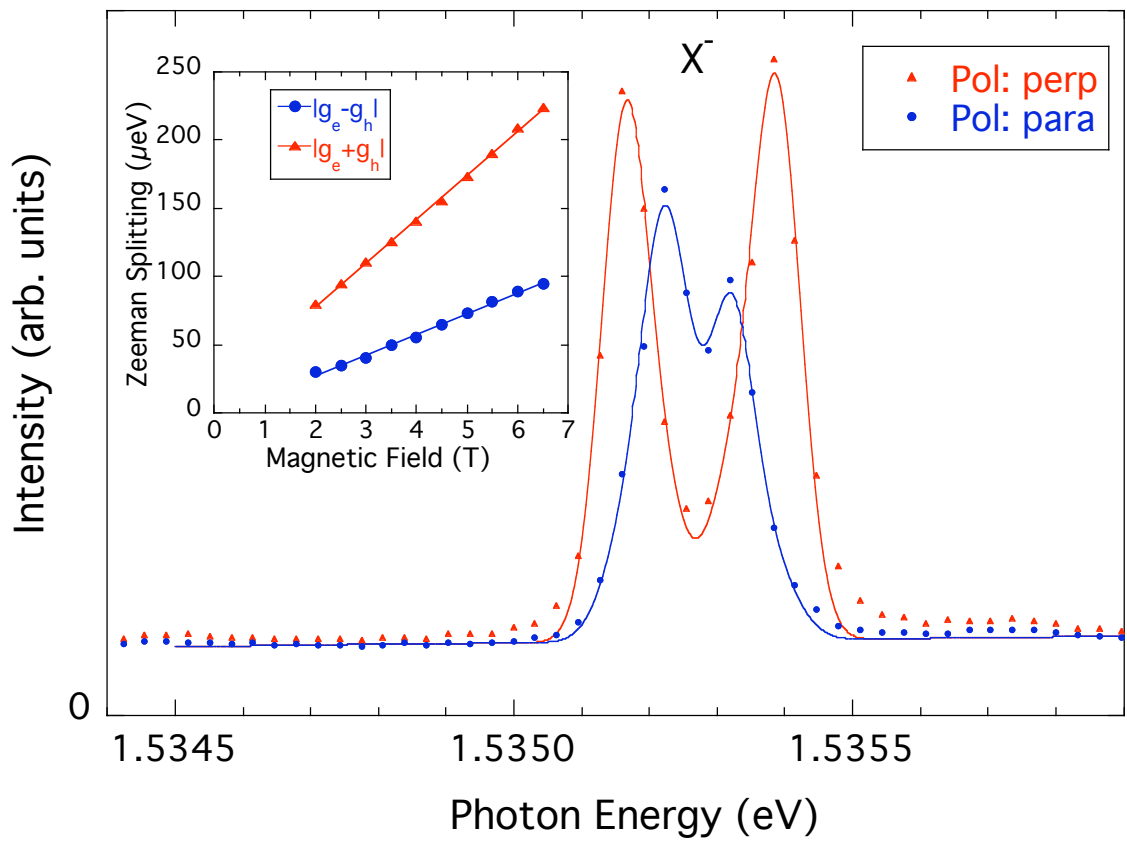

Figure 2

D.Y. Oberli

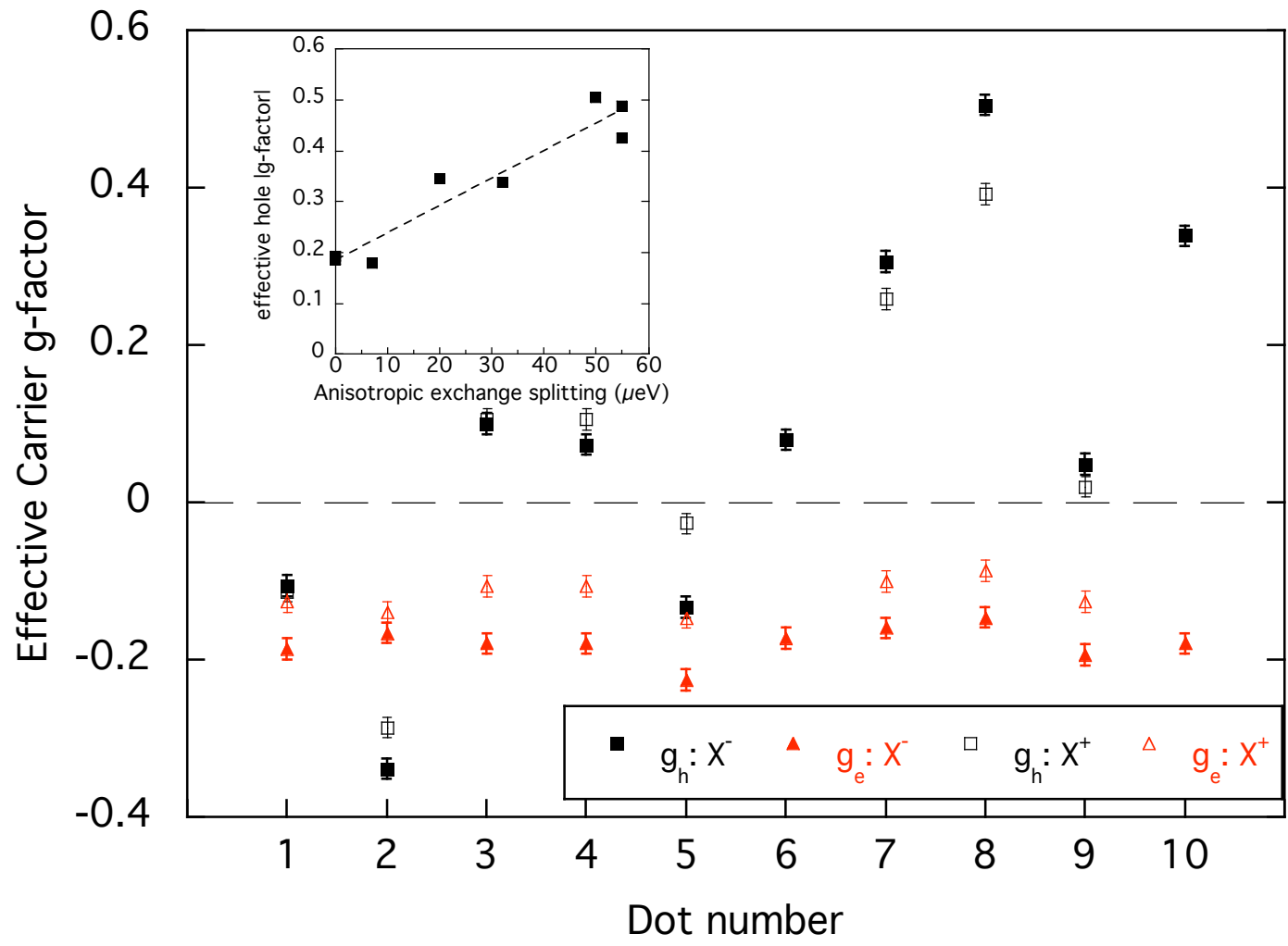

Figure 3

D.Y. Oberli